\documentclass{article}

\usepackage{xcolor,colortbl}
\definecolor{Gray}{gray}{0.95}
\newcolumntype{C}{>{\columncolor{Gray}}c}
\usepackage{color}
\usepackage{spconf,amsmath,graphicx}
\usepackage{hyperref}       % hyperlinks
\usepackage{url}            % simple URL typesetting
\usepackage{booktabs}       % professional-quality tables
\usepackage{amsfonts}       % blackboard math symbols
\usepackage{xcolor}         % colors
\usepackage{graphicx}
\usepackage{amsmath,amssymb}
\usepackage{xurl}
\usepackage{multirow}
\usepackage{tabularx}
\usepackage{booktabs}
\usepackage{makecell}
\usepackage{float}
\hypersetup{colorlinks=true}
\usepackage{caption}
\usepackage{subcaption}
\usepackage{cite}
\usepackage{tabularx}
\usepackage{booktabs}
\usepackage{adjustbox}
\usepackage{amsmath}
\usepackage{svg}
\usepackage{amssymb}

\title{Speech-Copilot: Leveraging Large Language Models for Speech Processing via Task Decomposition, Modularization, and Program Generation}
\name{Chun-Yi Kuan\textsuperscript{$\heartsuit$}$^{1}$, Chih-Kai Yang\textsuperscript{$\heartsuit$}$^{2}$, Wei-Ping Huang\textsuperscript{$\clubsuit$}$^{1}$, Ke-Han Lu\textsuperscript{$\clubsuit$}$^{1}$\thanks{\textsuperscript{$\heartsuit$}\textsuperscript{$\clubsuit$} Equal Contribution.}, Hung-yi Lee$^{1}$}

\address{\begin{tabular}[c]{@{}c@{}}
     $^{1}$Graduate Institute of Communication Engineering, National Taiwan University, Taiwan,\\
     $^{2}$Graduate Institute of Electrical Engineering, National Taiwan University, Taiwan
\end{tabular}}

\begin{document}
\ninept

\maketitle
\begin{abstract}

In this work, we introduce Speech-Copilot, a modular framework for instruction-oriented speech-processing tasks that minimizes human effort in toolset construction. Unlike end-to-end methods using large audio-language models, Speech-Copilot builds speech processing-specific toolsets by analyzing pre-collected task instructions and breaking tasks into manageable sub-tasks. It features a flexible agent that performs tasks through program generation based on large language models. Our approach achieves state-of-the-art performance on the Dynamic-SUPERB benchmark, demonstrating its effectiveness across diverse speech-processing tasks. Key contributions include: 1) developing an innovative framework for speech processing-specific toolset construction, 2) establishing a high-performing agent based on large language models, and 3) offering a new perspective on addressing challenging instruction-oriented speech-processing tasks. Without additional training required by end-to-end approaches, our method provides a flexible and extendable solution for a wide range of speech-processing applications.

\end{abstract}
%
% \vspace{-10pt}
\section{Introduction}
\label{sec:intro}

Nowadays, large language models (LLMs) have impacted the AI research community with their strong capabilities across a wide variety of natural language processing (NLP) tasks involving complicated reasoning~\cite{wei2022chain, zoph2022emergent, kojima2022large}, planning~\cite{huang2022language, huang2024understanding, valmeekam2023planning}, and self-reflection~\cite{huang2022large, pan2024automatically, miao2023selfcheck}. These extraordinary abilities have established LLMs as powerful tools for humans and have cemented their pivotal role in recent AI research.

Especially, the potential of employing LLMs as an ``assistant" or an ``agent"~\cite{schoenegger2024ai, xie2024large, tang2024prioritizing} has been extensively explored, with several LLM-based agents that can use tools, e.g. API calls, to solve tasks across diverse domains and various modalities~\cite{schick2023toolformer, suris2023vipergpt, huang2023audiogpt, du2024anytool} being proposed recently. However, we notice that: 1) Most of the agents in prior works rely on the pre-existing toolsets, which require significant manual efforts to collect and maintain. Only a few works~\cite{cai2023large, yuan2024craft, qian2023creator} explore the toolset construction process of LLM-based agents. 2) The development for speech-processing agents remains limited, restricting broader and more convenient applications of speech-processing technologies. These motivate us to start with a systematic methodology for speech-processing toolset construction that goes beyond human brainstorming and develop an LLM-based agent for speech-processing applications.

% While featuring the capabilities of reasoning, planning, and tool utilization of LLMs, these agents typically relied on pre-existing human-prepared toolsets, making their performances highly limited by the coverage and diversity of the toolsets, which are unfortunately inherently constrained by human cognition. Additionally, constructing high-quality toolsets with broad coverage and diversity requires substantial human effort. Thus, efficiently developing effective toolsets is crucial for creating powerful LLM-based agents and should be approached cautiously when developing new agents. 
%However, most prior works have not emphasized this aspect, 

% \input{images/image}

In this paper, we introduce \textbf{Speech-Copilot}, a general framework consisting of two main components: 1) a toolset construction method leveraging LLMs with minimal human efforts, and 2) an LLM-based agent serving as a scalable, interpretable, and flexible interface capable of solving a wide variety of speech-processing tasks via program generation. 

For the toolset construction, we propose a pipeline employing an LLM to analyze a diverse set of pre-collected task instructions that can be either collected from humans or synthesized by LLMs, identify the corresponding speech-processing tasks, and decompose these tasks into sub-tasks. This results in a set of unique and basic sub-tasks, which are subsequently formulated as code modules by LLMs and implemented by humans with suitable speech models. This approach enables near-automatic toolset development, significantly reducing the required manual effort while ensuring effectiveness and avoiding redundancy. Additionally, it is quite flexible and scalable, as users can freely choose the speech models they prefer for each module or add new modules if necessary. 

An LLM-based agent capable of utilizing these modules via programming is developed. Our results show that this agent can solve various tasks by appropriately combining the basic modules, achieving state-of-the-art performance on Dynamic-SUPERB~\cite{huang2024dynamic} compared with baselines including large audio-language models and cascaded systems. 
% as illustrated in Fig. \ref{fig:performance}. 
This validates the efficacy of Speech-Copilot. We also find that Speech-Copilot has strong multi-task ability that can deal with several tasks in a single user query without sacrificing the performances.
The demo page and code for Speech-Copilot are available\footnote{\url{https://github.com/kuan2jiu99/speech-copilot}\label{demo}}.
% A demo page is available\footnote{\scriptsize{Demo at: \url{https://sites.google.com/view/slt2024-demo-page}}\label{demo}}. 
% After review, we will release all the related code for Speech-Copilot for the community to use, hoping to make it more convenient for everyone to perform speech-processing tasks.
% We are releasing Speech-Copilot\footnote{\scriptsize{Demo at: \url{https://sites.google.com/view/slt2024-demo-page}}\label{demo}} for the community to use, hoping to make it more convenient for everyone to perform speech-processing tasks.
% Overall, our contributions are three-fold:
Overall, our contributions are:
\begin{enumerate}
    \item{Proposing a new toolset construction framework for LLM-based agents that requires minimal human efforts only.}
    
    \item{Building up a new speech-processing agent with LLMs, which achieves impressive benchmark performances.}
    
    \item{Releasing the agent as a public speech-processing toolkit.}
\end{enumerate}
% Overall, our contributions are:
% (1) Proposing a new toolset construction framework for LLM-based agents that requires minimal human efforts only, (2) Building up a new speech processing interface with LLMs, which achieves impressive benchmark performances, and (3) Releasing the agent as a public speech processing toolkit to the community.

\begin{figure*}[ht]
    \centering
    % \includesvg[width=1.0\textwidth]{images/images/speech-copilot-pipeline.svg}
    \includegraphics[width=\textwidth]{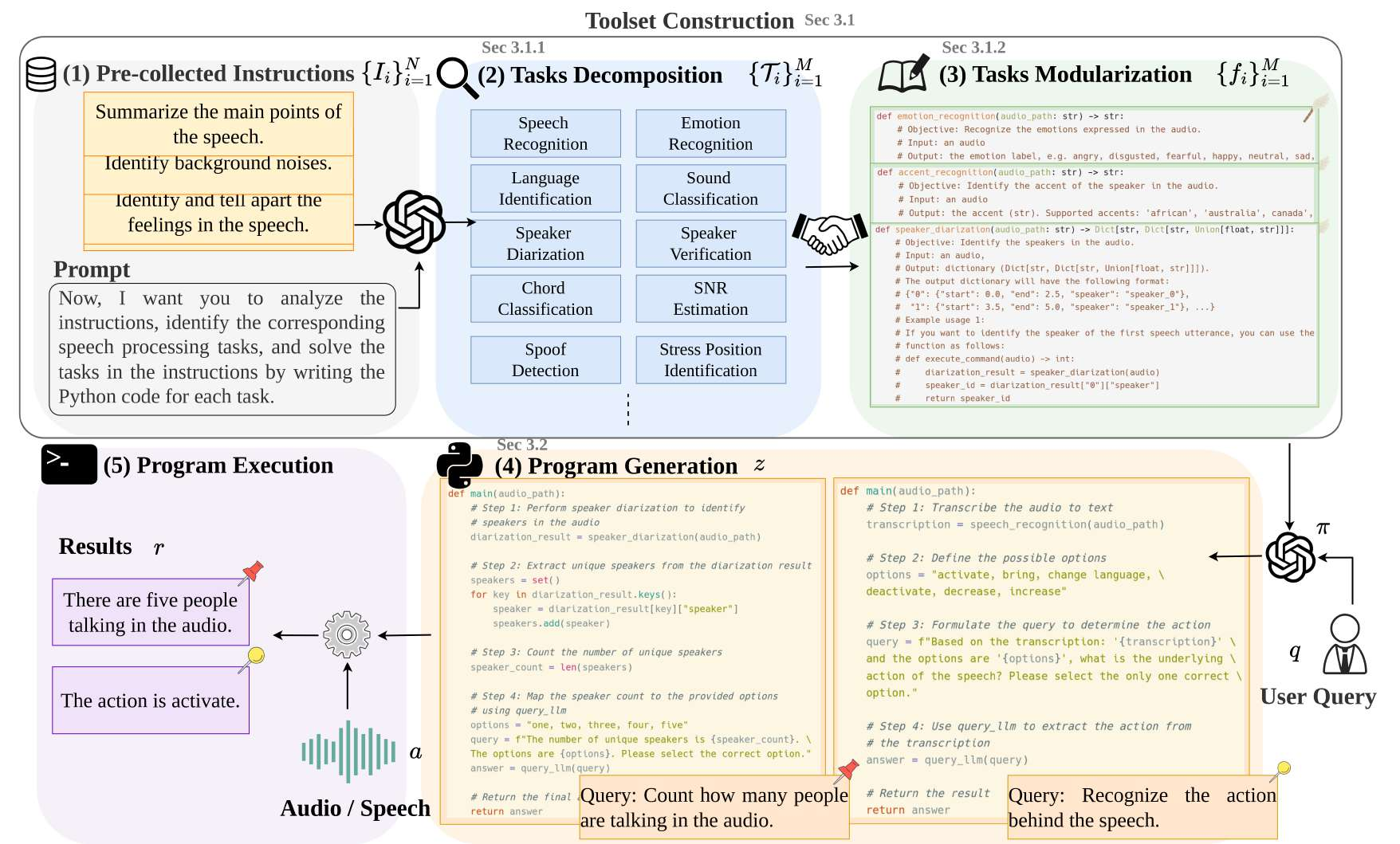}
    % \caption{Overview of Speech-Copilot. This diagram illustrates the three primary phases of Speech-Copilot: task decomposition, task modularization, and program generation. Each phase builds upon the previous one, starting with decomposing diverse speech-processing tasks into fundamental sub-tasks, followed by transforming these sub-tasks into detailed, documented modules, and concluding with implementing these modules with scientifically supported speech models and generating programs using the modules to solve the encountered tasks.}
    \caption{Overview of Speech-Copilot with the toolset construction and the program generation phases. During the toolset construction, we first conduct task decomposition to decompose diverse speech-processing task instructions into fundamental sub-tasks. Next, task modularization is performed to transform the sub-tasks into documented modules with LLM, manually implemented with scientifically grounded models. Finally, in the program generation phase, programs are generated by LLM based on the user query and executed on the audio input to get the result.
    Please refer to the demo page\footref{demo} for more details about prompts.}
    \vspace{-10pt}
    \label{fig:pipeline}
\end{figure*}
\vspace{-10pt}
\section{Related works}
\label{sec:related_works}

\subsection{Tool utilization of LLMs}

Large language models (LLMs) have been proven to be highly effective in many natural language processing (NLP) tasks. By integrating external tools, LLMs can enhance their functionality and handle a wider range of tasks using the additional knowledge and capabilities~\cite{huang2024planning, guo2024stabletoolbench, li2023api, yuan2024easytool, ye2024rotbench, shi2024learning, patil2023gorilla, qin2023toolllm, schick2024toolformer}. 
For example, AnyTool~\cite{du2024anytool} is an LLM-based agent that uses various APIs to answer user queries across different domains, such as providing specific information about a book, movie, or product, or offering personalized recommendations through connected recommendation APIs. 
Similarly, ViperGPT~\cite{suris2023vipergpt} employs LLMs for complex visual tasks by generating Python programs that coordinate vision-and-language models to process visual queries.

However, despite the advancements in NLP and computer vision, the exploration of using LLMs in the speech domain to integrate various speech modules and foundation models, particularly via program generation for speech/audio tasks, is relatively limited.

% Moreover, while LLM-based agents have been developing rapidly, the construction of toolsets for these agents remains relatively underexplored. Existing instance-level toolset creation approaches~\cite{cai2023large, yuan2024craft, qian2023creator} involve iteratively sampling problem-answer pairs from a dataset, selecting diverse pairs, generating solutions, and refining the toolset by removing incorrect solutions.
Moreover, while LLM-based agents have been developing rapidly, the construction of toolsets for these agents remains relatively underexplored. Existing approaches~\cite{cai2023large, yuan2024craft, qian2023creator} typically create tools at the instance level, where a new tool is created tailored for a single or a few instances, overlooking the high-level similarity of the collected instances. This may introduce redundancy in the created toolset that may be difficult to de-duplicate. In addition, they typically require golden labels for the instances, which make it harder to collect instances.

In contrast, our proposed method takes a holistic approach, utilizing LLMs to identify essential sub-tasks from all collected instructions simultaneously. This approach significantly reduces the redundancy of the constructed toolsets compared to instance-level creation methods. Additionally, since our method only requires instructions that can be easily synthesized, the associated audio files and golden labels are not necessary, making it more data-efficient and simplifying the toolset construction process compared to existing methods.

% Large language models (LLMs) have been proven highly effective across numerous natural language processing (NLP) tasks. 
% Their capacity to integrate external tools to enhance their functionality for practical challenges has drawn significant attention. 
% By equipping LLMs with tools, they can access a broader spectrum of tasks using the added knowledge and functionalities~\cite{huang2024planning, guo2024stabletoolbench, li2023api, yuan2024easytool, ye2024rotbench, shi2024learning}. 
% For instance, AnyTool~\cite{du2024anytool} is an LLM-based agent that harnesses a variety of APIs to respond to user queries across multiple domains. 
% These queries include specific information requests, e.g., details on a book, movie, or product, or recommendations tailored to user preferences through connected recommendation APIs. 
% Similarly, ViperGPT~\cite{suris2023vipergpt} utilizes LLMs for complex visual tasks, generating Python program that orchestrates vision-and-language models to process visual inquiries. 

% However, despite advancements in NLP and computer vision, the speech domain has seen limited exploration in using LLMs to integrate various speech modules and foundation models, particularly through program generation for speech and audio tasks.

\subsection{Toolkit Applications in Speech Processing}
Equipping LLMs with speech-processing toolkits is underexplored compared with NLP and computer vision domains. There is only a limited number of studies in this area. Among them, AudioGPT~\cite{huang2023audiogpt} is notable for using an LLM as a core controller to manage various pre-trained audio and speech models. Upon receiving a user query, AudioGPT analyzes it, classifies it into task families, and assigns an appropriate speech model for the task. The model's output is then sent back to the user as the system's response. However, AudioGPT has some limitations: \textbf{1) Limited generalizability}: It assigns only one model per query, with no collaboration between speech models, which limits its ability to handle complex tasks beyond predefined task families. Additionally, the families are not sufficiently broad and diverse, further restricting the generalizability. For instance, AudioGPT can not deal with Dynamic-SUPERB tasks, which are too complex to be addressed by a single model from its task families and model collection.
\textbf{2) Lack of flexibility}: The response is generated and provided to the user in a black-box manner, disallowing the users to manipulate the model's behavior and limiting the flexibility.

In contrast, Speech-Copilot effectively addresses these issues through a well-crafted toolset construction and program generation approach, where a toolset with fundamental speech modules is first constructed and the agent can then use these modules as basic blocks and dynamically combine them to solve various tasks based on the user query. This ensures the generalizability and versatility. By solving tasks through programming, Speech-Copilot allows users to modify the program according to their preferences, providing a higher degree of flexibility and enabling behavior manipulation to some extent. Furthermore, analyzing the agent's reasoning steps in the programs enhances interpretability, making it easier for users to understand how the solutions are derived. 

\section{Methods}
\label{sec:methods}
%In this section, we will introduce how Speech-Copilot is developed step by step.  We will explain how to break down a diverse set of pre-collected user instructions to create comprehensive toolsets. By combining the speech modules within these toolsets, we can form executable code that addresses the user's problems and needs. Additionally, we will introduce the speech modules we have selected.
The development of Speech-Copilot consists of two phases. The first phase is the toolset construction, in which an LLM is employed first to figure out the underlying common components of the pre-collected task instructions and then modularize the identified components into speech-processing modules. The other phase is program generation, which develops an agent solving various speech-processing tasks by writing a program to utilize the speech-processing modules. The overview of Speech-Copilot is illustrated in Fig. \ref{fig:pipeline}, and we detail the phases in Sec. \ref{construction} and \ref{development}.

\subsection{Toolset Construction}
\label{construction}

The toolset construction phase involves two steps: task decomposition and task modularization.

\subsubsection{Task Decomposition}
\label{decomposition}
Task decomposition aims to find out the common components, i.e. sub-tasks, of a wide variety of speech-processing tasks. We start from a set of diverse task instructions, which can be collected from real humans or synthesized with LLMs. Given a set of $N$ distinct instructions $\{ I_i \}_{i=1}^N$ that corresponds to $N$ different speech-processing tasks $\{ T_i \}_{i=1}^N$, the objective of task decomposition is to construct a set of $M$ sub-tasks $\{ \mathcal{T}_i \}_{i=1}^M$ such that each task $T_i$ can be represented as the combination of some sub-tasks, indexed by $J$ $\subseteq$ $\{1, 2,\ldots, M\}$, through a suitable combination function $h_i$
% \begin{align}
% \label{decomposition_constraint}
%     \forall T_i, i=1,2,\ldots,N, \exists h_i,  \text{ s.t. } T_i = h_i(\mathcal{T}_1, \mathcal{T}_2,\ldots,\mathcal{T}_M)
% \end{align}
\begin{equation}
\begin{aligned}
\label{decomposition_constraint}
    \forall T_i, \, i \in \{1, 2, \ldots, N\}, \quad &\exists J \subseteq \{1, 2, \ldots, M\} \text{ and } h_i, \\
    &\text{such that} \quad T_i = h_i\left(\{\mathcal{T}_j \mid j \in J\}\right)
\end{aligned}
\end{equation}
Here, the combination of ``sub-tasks" means solving the sub-tasks and combining the corresponding results to solve the target task. For example, the speech-to-text translation (ST) can be solved by first conducting automatic speech recognition (ASR) and passing the transcription to a text-based translation model. Hence we say that the ST task is the combination of ASR and text-based translation, with simple cascading as the combination function.

We adopt an LLM for task decomposition. The employed LLM is first asked to map each instruction $I_i$ to the task $T_i$ by prompting it to analyze the instruction and identify the corresponding speech-processing task through chain-of-thought reasoning~\cite{wei2022chain}. As for the decomposition, besides the requirement in Eq. \ref{decomposition_constraint}, it is desirable for the constructed sub-tasks to be fundamental enough to remain useful and transferable for unseen tasks, thereby ensuring the constructed set is compact. To achieve this, instead of using common toolset construction methods that create tools based on single instructions~\cite{qian2023creator}, we require the LLM to consider all instructions simultaneously. This approach allows the LLM to identify common components across different tasks and decompose them into sub-tasks that are either shared among several tasks or unique to a single task. De-duplication of sub-tasks is also conducted with self-reflection~\cite{huang2022large} to further reduce the redundancy. Finally, a set of sub-tasks $\{ \mathcal{T}_i \}_{i=1}^M$ with low redundancy can be constructed, where the intention of the sub-tasks can be interpreted with the reasoning in LLM's response.

\subsubsection{Task Modularization}
\label{modularization}
The task modularization involves transforming the constructed sub-tasks $\{ \mathcal{T}_i \}_{i=1}^M$ from the previous task decomposition step into a set of modules $\{f_i \}_{i=1}^M$ with an LLM, and implementing those modules with existing speech-processing models. Specifically, as the modules involve solving complex speech-processing tasks rather than simple algorithmic problems, it is not required for the LLM to implement the modules directly. Instead, the modularization requires the LLM to consider all the sub-tasks and formulate each sub-task as a module, where a module is defined to be a code block with detailed documentation that should include (1) the module name, (2) the overall objective of the module, i.e. the task information associated with the module, (3) the input to the module as well as the data type and format, (4) the output of the module along with the data type and format, and (5) example usages of the module demonstrating how it can be used individually or together with other modules. 

With in-context learning, the LLM is capable of generating high-quality documentation. This reduces the required efforts of human developers, and the developers can freely modify the documentation according to their preferences or for the convenience of implementation. The constructed modules $\{f_i \}_{i=1}^M$ will serve as part of the input to the LLM-based agents, as detailed in Sec. \ref{development}.

% \subsubsection{Module Implementation}
% \label{module_implement}
The modules are manually implemented with suitable speech models based on the objective of the modules. We adhere to some principles when selecting the models: \textbf{1) Scientific support:} Only the models with publicly available papers or technical reports will be included. This ensures the reliability of the selected models and their sources. \textbf{2) Clear guidelines:} As we plan to publicly release Speech-Copilot to the community, the selected models should be accompanied by clear guidelines on how to use them, e.g. the environment setup, the running command, etc, ensuring user-friendliness. 

These principles ensure that the speech models included in Speech-Copilot are not only scientifically robust but also practical and easy to use, facilitating widespread adoption and application.

\subsection{Program Generation}
\label{development}
During the program generation phase, an LLM-based program-generating agent $\pi$ is constructed. Given a textual query $q$ and audio input $a$ from the user that is unseen during the toolset construction phase, the objective of the agent $\pi$ is to generate a program $z$ $=$ $\pi(q | \{f_i \}_{i=1}^M)$ based on the available modules $\{f_i \}_{i=1}^M$ from Sec. \ref{modularization} that solves the target task $T$ specified by the query.

This process involves the following steps: (1) Identifying the task $T$ from the query $q$, (2) Determining the combination $h$ and a set of sub-tasks $\mathcal{T}_q$ $\subseteq$ $\{ \mathcal{T}_i \}_{i=1}^M$ that satisfy $T$ $=$ $h(\mathcal{T}_q)$, (3) Selecting the relevant modules $f_q$ $=$ $\{f_i|\mathcal{T}_i \in \mathcal{T}_q, i\in \{1, 2, \ldots, M\} \}$, and (4) Generating a program $z$ that utilizes the modules $f_q$ and integrates their results with the combination $h$. In practice, we develop the agent $\pi$ by guiding the LLM through these steps, providing the module documentation $\{f_i \}_{i=1}^M$, and specifying additional constraints in the prompt. These constraints include requiring the agent to provide reasoning and explanations for how it addresses each step and to generate the program $z$ in a specific format, enhancing the interpretability and ease of parsing the programs generated by the agent.  Finally, the result $r$ of the query $q$ can be obtained by executing the program $z$ on the audio input $a$ with a Python interpreter.

\section{Experimental Setups}
\label{sec:evaluation}
\subsection{Evaluation Benchmark}
We evaluate Speech-Copilot on Dynamic-SUPERB\footnote{\label{fn:official-page}{\scriptsize\url{https://dynamic-superb.github.io/}}} \cite{huang2024dynamic} because of its wide coverage of diverse, complex, and challenging speech and audio tasks. 
Dynamic-SUPERB is designed to assess universal speech models that perform diverse and complex tasks with both strong instruction-following abilities and complicated speech/audio-related understanding. 
It includes 55 tasks, involving complicated speech/audio-related reasoning and covering six aspects of the speech and audio modalities: audio, content, degradation, semantic, paralinguistic, and speaker. The aspects are explained as follows:

\begin{enumerate}
    \item {\textbf{Audio (AUD)}: 
    This aspect assesses the model's ability to interpret audio signals. Tasks in this aspect include detecting the sound of a specific object in an audio clip and classifying the sound into some categories.}

    \item {\textbf{Content (CNT)}: 
    This aspect evaluates the model's understanding of speech content. Tasks involve speech command identification, language recognition, and so on.}
    
    \item {\textbf{Degradation (DEG)}: 
    The goal here is to measure the model's ability to detect noise and reverberation in speech signals. Tasks include predicting the signal-to-noise ratio (SNR) at the utterance level and determining if speech signals are affected by reverberation.}
    
    \item {\textbf{Paralinguistic (PRL)}: 
    The tasks of this aspect aim to gauge the model's understanding and the ability to make inferences based on paralinguistic information in speech. The tasks include but are not limited to accent identification, emotion recognition, etc.}
    
    \item {\textbf{Semantic (SEM)}: 
    The goal is to assess the model's semantic understanding. 
    This aspect involves intent classification, sarcasm detection, etc.}
    
    \item {\textbf{Speaker (SPK)}: 
    The aim is to evaluate the model's capacity to extract speaker-related information. 
    This involves tasks like speaker verification and multi-speaker detection.}
    
\end{enumerate}

Currently, Dynamic-SUPERB tasks are formulated as multiple-choice questions with one and only one golden label per question.

\subsection{Metrics}
Dynamic-SUPERB tasks are evaluated in accuracy, where a hit occurs when the response of the evaluated model aligns with the golden label. Due to the generative nature of current models, e.g. LLMs or large audio-language models (LALMs) in Sec.~\ref{baseline}, instead of a single option, the models tend to generate long-form responses that do not follow a certain format, making the conventional exact-match (EM) evaluation unsuitable. To this end, the evaluation policy of Dynamic-SUPERB uses an LLM, whose ability as the automatic evaluator has been studied~\cite{chiang2023can, chiang2023closer}, to determine the alignment between the models' predictions and the labels.
% Previous studies~\cite{chiang2023can, chiang2023closer} have shown that LLMs can be used for automatic evaluation.

In this work, GPT-4o\footnote{\label{fn:gpt4_version}{\scriptsize{The model version employed in this study is gpt-4o-2024-05-13.}}} is used as the evaluator.  
% When given a sample and evaluation criteria, LLMs can generate results that closely match those of human experts in the field.
During the grading process, the original task instructions, the response from the assessed models, and the golden labels are presented in the prompts to the evaluator. 
We also include some rules in the prompts that the evaluator should strictly follow, and the evaluator is then employed for judgment. The rules include:
\begin{enumerate}
    \item As each question in Dynamic-SUPERB has one and only one golden answer, it should be judged \textbf{incorrect} if the models choose \textbf{no or multiple} options, meaning that they should clearly select one and only one option for each question.
    
    \item The evaluator should provide reasons for their judgment. 
    
    \item The evaluation should be summarized with a single "Yes/No" in a specific format, where "Yes" indicates that the prediction aligns with the label, and "No" indicates it does not.
\end{enumerate}
Besides automatic evaluation with GPT-4o, human verification of the evaluation results is conducted to ensure grading correctness and consistency. Finally, we follow the standard approach of Dynamic-SUPERB and report the average performances of the aspects.
\vspace{-5pt}

% Since our selected baseline models do not follow a specific format for responses to instructions in different tasks, we use GPT-4o as the grading model to ensure fairness and impartiality in the grading process.
% We compare the responses from large audio-language models with the corresponding ground truth labels using GPT-4. 
% The grading method involves determining whether the response's meaning aligns with the ground truth label. 
% The grading model outputs either "Yes" or "No." "Yes" indicates that the response and the ground truth label have aligned meanings, while "No" indicates they do not. 

% For example, in a sound classification task, the instruction is ``We require you to categorize the sounds that belong to the environment. The answer could be dog, rooster, pig, cow, frog, cat, hen, insects, sheep, or crow''.
% If the large audio-language model's response is ``The sound is of a cat meowing'', and the ground truth label is ``cat'', the grading model will determine that the response aligns with the ground truth, marking it as correct.
% Due to the inherent randomness in GPT-4o grading, we ran the grading process three times and took the average.

\subsection{Baselines}
\label{baseline}
We compare the performances of Speech-Copilot with those of several baseline models to verify the effectiveness of Speech-Copilot, including the toolset construction and program generation. 
The baselines include several recent publicly available large audio language models (LALMs), e.g.  Qwen-Audio-Chat~\cite{chu2023qwen}, SALMONN~\cite{tang2023salmonn}, LTU-AS~\cite{gong_ltuas} and WavLLM~\cite{hu2024wavllm}, and cascaded systems that employ LLM to solve the tasks based on the information from automatic speech recognition (ASR), automatic audio captioning (AAC), and other available speech models. Large audio-language models~\cite{lu2024destaenhancingspeechlanguage, chu2023qwen, tang2023salmonn, gong_ltuas, hu2024wavllm, kong2024audio, gong2023listen} extend the capabilities of standard large language models by incorporating audio and speech recognition features. This integration enables LALMs to process and respond to tasks involving sound and speech.

On the other hand, in the cascaded systems, we adopt GPT-3.5\footnote{\label{fn:gpt3.5_version}{\scriptsize{The model version employed in this study is gpt-3.5-turbo-0125.}}}~\cite{ChatGPT} as the LLM and  Whisper-large-v3~\cite{radford2023robust} and Qwen-Audio-Chat~\cite{chu2023qwen} for ASR and AAC, respectively. 
The cascaded systems are denoted as ``ASR+LLM" and ``ASR+AAC+LLM" for those using ASR results only and using both ASR and AAC simultaneously. We also compare Speech-Copilot with the cascaded system that provides all the information from our constructed modules listed in Sec. \ref{sec:toolset}, except for the speaker verification module if there's only one audio input, to the LLM, denoted as ``All Attributes + LLM". This baseline simulates the ultimate cascaded system where the LLM utilizes all the available information to make predictions. 

% In addition, we compare Speech-Copilot with the cascaded system that provides all the information from our constructed modules listed in Sec. \ref{sec:toolset}, except for the speaker verification module if there's only one audio input, to the LLM, denoted as ``All Attributes + LLM". This baseline simulates the ultimate cascaded system where the LLM utilizes all the available information to make predictions. 

% \vspace{-8pt}
\subsection{Setup}
\label{setup}
Greedy decoding is applied to all the models in our experiments. Regarding the candidates of LLMs, in the stages of toolset construction, program generation, and evaluation, GPT-4o is adopted because of its strong language capabilities. On the other hand, when executing the generated programs, GPT-3.5 is employed if querying LLM is required for certain modules. This choice balances cost and model performance, i.e. we can use a more powerful model where it's most needed while opting for a less expensive option elsewhere. 

As for Whisper-large-v3, though prompting for Whisper is common in prior works~\cite{concat, yang2023investigating, zhuo2023lyricwhiz}, it is not employed in our experiments due to the unclear effect of prompting methods of Whisper~\cite{yang2024prompts}. For other models involved in this work, we used the default settings, with the generation strategy consistently set to greedy decoding.

% We select four publicly available large audio-language models (LALMs) and the cascaded pipeline of automatic speech recognition and LLM as our baselines. 
% Large audio-language models include Qwen-Audio-Chat~\cite{chu2023qwen}, SALMONN~\cite{tang2023salmonn}, LTU-AS~\cite{gong_ltuas} and WavLLM~\cite{hu2024wavllm}. 
% The cascaded pipeline means using Whisper-large-v3~\cite{radford2023robust} to get the speech recognition results from the input audio first, then feeding it along with task instructions to gpt-3.5-turbo~\cite{ChatGPT}.
\section{Results}
\label{sec:results}
\subsection{Toolset Construction}
\begin{table*}[th]
  \caption{Accuracy (\%) of the models across the aspects of Dynamic-SUPERB.
  The best performance in each aspect is marked in bold, while the second-best one is underlined.
  ``\# of Tasks" represents the number of tasks under each aspect in Dynamic-SUPERB.}
  \label{tab:dynamic-superb-main-results}
  \footnotesize
  \setlength{\tabcolsep}{5.3pt}
  \centering
  \vspace{-5pt}
  \begin{tabular}{ l | c  c  c  c  c  c | c}
    \toprule
    % \multirow{2}{*}{\textbf{Model}} & & & & & & & \multirow{2}{*}{} \\
     & {\textbf{Audio}} 
     & \textbf{Content} 
     & \textbf{Degradation} 
     & \textbf{Paralinguistics} 
     & \textbf{Semantic} 
     & \textbf{Speaker} 
     & \textbf{Average} \\
    \# of Tasks & 7 & 11 & 19 & 7 & 6 & 5 & 55 \\
    \midrule

    Qwen-Audio-Chat~\cite{chu2023qwen} 
    & \underline{73.2} 
    & 63.3 
    & 31.1 
    & 29.3
    & 48.1 
    & 41.4 
    & 45.5 \\

    SALMONN~\cite{tang2023salmonn}
    & 15.0 
    & 52.0 
    & 28.2 
    & 24.5 
    & 50.8 
    & 33.2 
    & 33.7 \\
    
    LTU-AS~\cite{gong_ltuas}
    & 14.5 
    & 44.0 
    & 37.5
    & 17.1 
    & 36.0 
    & 40.2 
    & 33.4 \\

    WavLLM~\cite{hu2024wavllm}
    & 22.3 
    & 53.3 
    & 36.8 
    & 24.6 
    & 51.0 
    & 22.3 
    & 36.9 \\
    \midrule

    ASR + LLM
    & 9.6 
    & 74.4
    & 44.6
    & \underline{33.1}
    & \underline{71.5}
    & 42.5
    & 47.4 \\

    ASR + AAC + LLM
    & 60.7
    & \underline{81.6}
    & {48.9} 
    & 32.6 
    & \textbf{72.8}
    & {46.4} 
    & {57.3} \\

    All Attributes + LLM
    & 62.4
    & {70.7}
    & \underline{56.8} 
    & 30.6
    & 68.5
    & \underline{62.5}
    & \underline{58.7} \\
    \midrule

    \textbf{Speech-Copilot (Ours)}
    & \textbf{73.4}
    & \textbf{90.7}
    & \textbf{{64.3}}
    & \textbf{56.6}
    & {70.7}
    & \textbf{{86.1}}
    & \textbf{72.4} \\

    % \textbf{Speech-Copilot (Ours)}
    % & \textbf{75.5}
    % & \underline{87.6}
    % & \textbf{\underline{64.3}}
    % & \underline{56.1}
    % & {68.5}
    % & \textbf{\underline{86.1}}
    % & \underline{71.1} \\

    % \textbf{Speech-Copilot (Ours)}
    % & \textbf{76.2} 
    % & \underline{87.6}
    % & \textbf{64.3}
    % & \underline{56.1}
    % & 68.5
    % & \textbf{86.1} 
    % & \underline{73.0} \\
    
    \bottomrule
  \end{tabular}
\end{table*}

% \cellcolor{orange!50}
\label{sec:toolset}
We first demonstrate the results of our toolset construction phase. We compare our task decomposition method, where we take all the instructions into consideration at once, with instance-level toolset creation, in which the tools are created based on a single instruction. More than 50 task instructions generated by GPT-3.5 are used for the construction, and the results with GPT-4~\cite{openai2023gpt4} are shown in Table \ref{tab:decomposition}. 

\label{toolset_result}

\begin{table}[ht]
    \centering
    \small % 使用較小的字體
    %\begin{adjustbox}{max width=\textwidth} % 確保表格不會超出頁面的範圍
    \caption{Number of sub-tasks with different toolset creation methods. "w/ reflection" means self-reflection is used to de-duplicate similar sub-tasks, while "w/o reflection" means no reflection is used.}
    \vspace{-5pt}
    \resizebox{\columnwidth}{!}{
    \begin{tabular}{cccc}
        %\hline
        \toprule
         & \makecell[c]{\textbf{Ours} \\ \textbf{(w/ reflection)}} & \makecell[c]{Ours \\ (w/o reflection)} & \makecell[c]{Instance-level \\ creation}\\
        \midrule
        \makecell[c]{\# of sub-tasks ($\downarrow$)} & \textbf{16} & 18 & 25 \\
        \bottomrule
    \end{tabular}
    }
    %\end{adjustbox}
    \label{tab:decomposition}
    \vspace{-8pt}
\end{table}

\begin{table}[ht]
    \centering
    %\small % 使用較小的字體
    %\begin{adjustbox}{max width=\textwidth} % 確保表格不會超出頁面的範圍
    \caption{Selected speech/audio models used for various modules.}
    \resizebox{\columnwidth}{!}{
    \begin{tabular}{c|c}
        %\hline
        \textbf{Modules} & \textbf{Selected Model} \\
        \hline
        \hline
        Speech Recognition & Whisper-large-v3~\cite{radford2023robust} \\
        \hline
        Language Identification & Whisper-large-v3~\cite{radford2023robust} \\
        \hline
        Speech Detection & Qwen-Audio-Chat~\cite{chu2023qwen} \\
        \hline
        Speech Emotion Recognition &emotion2vec~\cite{ma2023emotion2vec} \\
        \hline
        Speech-to-Noise Ratio (SNR) Estimation & Brouhaha~\cite{lavechin2023brouhaha} \\
        \hline
        Reverberation Detection &  Qwen-Audio-Chat~\cite{chu2023qwen} \\
        \hline
        Accent Classification & CommonAccent~\cite{zuluaga2023commonaccent} \\
        \hline
        Stress Position Identification & Whisper-large-v3~\cite{radford2023robust}, GPT-3.5~\cite{ChatGPT} \\
        \hline
        Spoofing Detection & Qwen-Audio-Chat~\cite{chu2023qwen} \\
        \hline
        Music Chord Classification & autochord~\cite{bayron2021autochord} \\
        \hline
        Sythetic Speech Detection & Qwen-Audio-Chat~\cite{chu2023qwen} \\
        \hline
        Speaker Verification & NVIDIA TitaNet-Large~\cite{9746806, kuchaiev2019nemo} \\
        \hline
        Speaker Diarization & pyannote speaker-diarization-3.1 ~\cite{Plaquet23} \\
        \hline
        Sound Classification & Qwen-Audio-Chat~\cite{chu2023qwen}, GPT-3.5~\cite{ChatGPT} \\ 
        \hline
        Query LLM & GPT-3.5~\cite{ChatGPT} \\
        \hline
        Speaker Distance Estimation & Qwen-Audio-Chat~\cite{chu2023qwen} \\
        \bottomrule
    \end{tabular}
    }
    %\end{adjustbox}
    \label{tab:speech_models}
    \vspace{-8pt}
\end{table}

As expected, our decomposition method, which considers all instructions collectively, significantly reduces the size of the sub-task set compared to instance-level tool creation. This reduction facilitates subsequent modularization and implementation. In addition, the size can be further reduced if the LLM is required to reflect on whether there are similar sub-tasks that can be unified or combined into single speech-processing tasks. This encourages the LLM to unify tasks of similar nature. For instance, the LLM can combine tasks related to the number of speakers with speaker diarization sub-tasks, thereby reducing the number of required sub-tasks.

As for the task modularization, we again employ GPT-4o to transform the collected sub-tasks into modules with detailed documentation with requirements outlined in Sec. \ref{modularization}. The resulting modules are listed in Table \ref{tab:speech_models}. We select the models for those modules based on the principles clarified in Sec. \ref{modularization}. For some modules where suitable and publicly available models are unavailable, we employ Qwen-Audio-Chat as the foundation model and realize these modules by prompting Qwen-Audio-Chat with a fixed set of prompts. The selected models for the modules are listed in Table \ref{tab:speech_models}.

Regarding the implementation details, most of the modules can be realized with the selected models directly. We briefly explain the implementation for some modules that require special handling. Stress position identification, a module for identifying stress syllables in spoken words, lacks a specialized model designed for this module. We hypothesize that this task can be approximated by combining speech recognition and LLMs due to their strong linguistic knowledge~\cite{suvarna2024phonologybench}. Thus, we implement this module using Whisper-large-v3 and GPT-3.5. Sound classification is a module for classifying a wide variety of sounds, e.g. environmental sounds, animal sounds, etc. Due to the high diversity of sound categories, it is difficult to find a single model to handle all kinds of classification tasks, so we choose to use Qwen-Audio-Chat as the backbone. To maintain generalizability, instead of using specific prompts for certain kinds of sound classification, we require Qwen-Audio-Chat to generate detailed audio captions, from which GPT-3.5 is adopted to extract the desired information based on the specific task objective.
\vspace{-5pt}

\subsection{Benchmark Performances}
We then compare the performances of Speech-Copilot with the selected baselines on Dynamic-SUPERB, as shown in Table \ref{tab:dynamic-superb-main-results}. Overall, Speech-Copilot achieves the highest average score across the 55 tasks and outperforms other baselines in 5 out of 6 aspects, indicating the efficacy of the constructed toolset and the problem-solving capability of the LLM-based program-generating agent.

We observe that the LALM baselines, i.e. Qwen-Audio-Chat, SALMONN, LTU-AS, and WavLLM, typically encounter significant issues that impact their performances. For instance, Qwen-Audio-Chat suffers from severe hallucinations about the spoken content in the provided audio when required to answer the questions directly. However, it can almost correctly identify the spoken content if asked to provide an audio caption of the audio input instead. This observation aligns with prior works~\cite{kuan2024understanding} on the hallucination of LALMs, indicating the vulnerability of recent LALMs. We also notice that these models struggle with clearly selecting one and only one option for the questions. For example, SALMONN tends to list all of them in the response, which is unacceptable.
% We also notice that these LALMs underperform relative to the cascaded systems. Even the most naive ``ASR+LLM" system can outperform these end-to-end LALMs, showing that the current LALMs are not strong enough for complex and challenging tasks that involve complicated speech/audio understanding and reasoning, like those in Dynamic-SUPERB. This suggests there is still room for improvement in these end-to-end models. Compared with LALMs, Speech-Copilot shows strong robustness to these issues due to its modular design and achieves significantly better performance on the evaluation benchmark.
Moreover, these LALMs underperform compared to cascaded systems. Even the simplest ``ASR+LLM" systems outperform end-to-end LALMs, indicating that current LALMs are not yet capable of handling complex speech/audio understanding and reasoning tasks, such as those in Dynamic-SUPERB. In contrast, Speech-Copilot demonstrates strong robustness due to its modular design, achieving significantly better performance on the evaluation benchmark.

% Compared with them, Speech-Copilot shows strong robustness to these issues due to its modular design and achieves significantly better performance on the evaluation benchmark.
% It is noted that the cascaded baselines, even the simplest ``ASR+LLM" system, typically outperform the end-to-end LALMs, showing that the current LALMs may not be strong enough for complex tasks that involve complicated speech/audio understanding and reasoning. This suggests there is still room for improvement in these end-to-end models.

Comparing the performances between the cascaded systems, it seems that the overall performance will be better if more information from different speech models is provided to the LLM. However, they still underperform relative to Speech-Copilot.
% Speech-Copilot also achieves better average scores and outperforms the cascaded baselines in most aspects of Dynamic-SUPERB.
Notably, there is a significant performance gap between the ``All Attributes + LLM" baseline, which is the most effective cascaded baseline in terms of the average score, and Speech-Copilot, despite using the same modules. The key difference lies in the utilization of information from these modules. The former incorporates all available information indiscriminately, regardless of the query's purpose, while the latter selectively uses and combines relevant modules based on an analysis of the input queries. This highlights the importance of selecting related and useful information to avoid being misled by redundancy. Additionally, the computation cost of the former system is higher since it requires running all the modules for all data, showcasing the benefit of efficient information selection considering computation budgets. Such a selection process is achieved by query analysis and programming in Speech-Copilot, demonstrating the advantage of a program-generating agent.

% In sum, Speech-Copilot is validated to be effective and strong by superior benchmark performances. It is also robust against several issues like hallucinations and misleading redundant information, making it a good agent for speech-processing tasks.

\begin{table}[ht]
\setlength\tabcolsep{3 pt}
    \centering
    \small
    \caption{The average number of reasoning steps and modules used.}
    \vspace{-5pt}
    % \resizebox{\columnwidth}{!}{
        \begin{tabular}{c | c c c c c c | c}
        \toprule
        & \textbf{AUD} & \textbf{CNT} & \textbf{DEG} & \textbf{PRL} & \textbf{SEM} & \textbf{SPK} & \textbf{AVG} \\
        \midrule
        % \# of Steps 
        Avg \# of steps 
        & 4.7 %$\pm$ 1.3 
        & 3.6 %$\pm$ 0.6 
        & 3.8 %$\pm$ 0.6 
        & 3.8 %$\pm$ 0.8 
        & 4.3 %$\pm$ 1.0 
        & 3.1 %$\pm$ 1.0 
        & 3.9 \\ %$\pm$ 0.9 \\
        \midrule
        % Std & 1.3 & 0.6 & 0.6 & 0.8 & 1.0 & 1.0 & 0.9 \\
        Avg \# of modules
        & 2.0
        & 1.7
        & 1.4
        & 1.7
        & 2.1
        & 1.1
        & 1.6 \\
        \bottomrule
        
        \end{tabular}
    % }
    \label{tab:reasoning step}
\end{table}

\vspace{-17.5pt}
\subsection{Further Study}
\subsubsection{Statistics on Reasoning Steps and Modules Used}

% We analyzed the complexity of programs generated by Speech-Copilot by evaluating the number of reasoning steps and modules utilized. 
% Reasoning steps encompass operations such as module calls, conditional and iteration statements, critical for solving problems. 
% For example, Fig.~\ref{fig:multi task example} illustrates a program with five reasoning steps. 
% The count of reasoning steps reflects the program's complexity, showcasing the model's capacity to integrate modules for decision-making and to handle module outputs through conditional and iterative processes.
% Results presented in Table \ref{tab:reasoning step} from the Dynamic-SUPERB benchmark indicate that Speech-Copilot employs 3 to 5 reasoning steps to address tasks, demonstrating its proficiency in conducting complex operations like multi-module utilization and algorithm design, beyond merely selecting and outputting module results. 
% The disparity between the numbers of reasoning steps and modules used highlights our pipeline's ability to not only employ modules but also to execute essential reasoning with them, necessary for solving tasks that single modules alone cannot address.

We analyzed the complexity of the programs generated by Speech-Copilot by examining the number of reasoning steps and modules used. 
Reasoning steps include operations like module calls, conditional statements, iteration statements, etc., that are necessary for the LLM to solve problems. 
For instance, Fig.~\ref{fig:multi task example} shows 5 steps. 
The number of reasoning steps represents the overall complexity of the programs, considering the model's diverse behaviors, such as combining modules for decision-making or processing module outputs with
conditional and iteration operations.
Table \ref{tab:reasoning step} presents the results across aspects of Dynamic-SUPERB. 
Speech-Copilot takes 3 to 5 reasoning steps when programming to solve these tasks, indicating its ability to perform complicated operations, like utilizing multiple modules or designing algorithms, rather than simply selecting a module and returning its output. 
Furthermore, the difference between the number of reasoning steps and used modules shows that our proposed pipeline not only uses modules but also performs necessary reasoning based on those modules, as these tasks cannot be solved using a single module.

\vspace{-5pt}
\subsubsection{In-the-wild Multi-task Examples}

\begin{figure}[ht]
    \centering
    \includegraphics[width=0.5\textwidth]{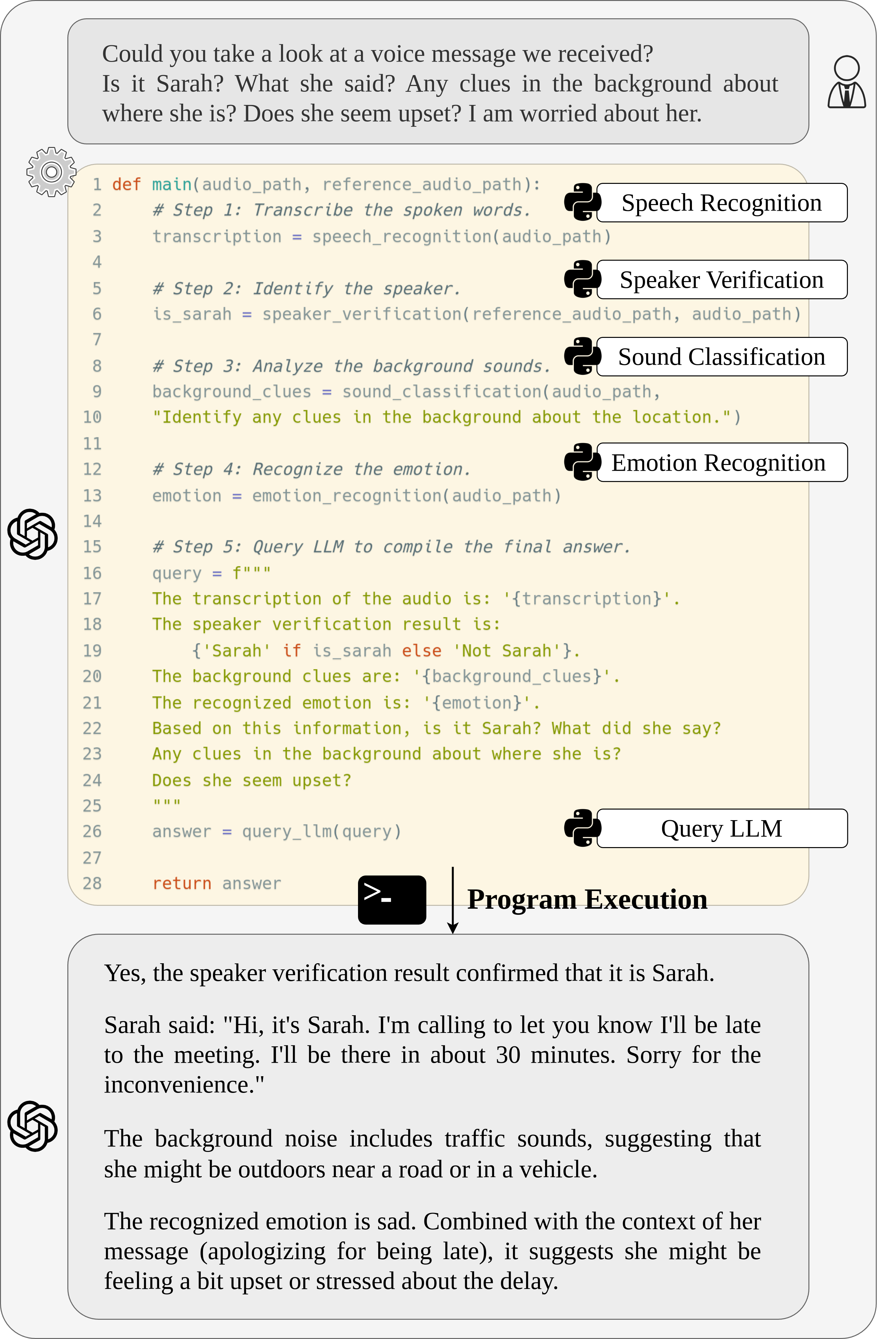}
    \caption{The results of Speech-Copilot on multi-task examples.}
    \label{fig:multi task example}
    \vspace{-10pt}
\end{figure}

Fig.~\ref{fig:multi task example} demonstrates a practical application of our pipeline through a real-life scenario, showing how various audio and speech-processing modules collaborate to provide a comprehensive understanding of a voice message. In contrast, existing large audio-language models, while proficient in speech recognition, often struggle with multitasking capabilities. 
These models face challenges in simultaneously processing speaker identity, emotion recognition, and sound classification while generating informative textual output.
This comparison highlights the advantages of our proposed pipeline over existing large audio-language models. 
By integrating multiple audio and speech-processing modules, our pipeline can deliver a more comprehensive and informative analysis of the given audio.
% \begin{figure}[ht]
%     \centering
%     \includesvg[width=0.5\textwidth]{images/images/demo.svg}
%     % \includegraphics[width=0.5\textwidth]{images/images/code.png}
%     \caption{Multi-task Examples.}
%     \label{fig:multi task example}
% \end{figure}

\vspace{-5pt}
\section{Conclusion and Future work}
\label{sec:conclusion}
\vspace{-5pt}
Speech-Copilot provides a practical and efficient approach to handling diverse, instruction-oriented speech-processing tasks. By breaking down complex instructions into manageable sub-tasks, formulating sub-tasks into code modules, and employing a flexible program-generating LLM-based agent to utilize the modules, our framework reduces the effort needed for toolset construction and enhances performance, achieving state-of-the-art performance on Dynamic-SUPERB benchmark. 
Our future work could explore using multiple modules or foundation models for each function and applying reinforcement learning from human feedback~\cite{ouyang2022training} to optimally select the best module for a given task. Furthermore, to address the evolving and diverse nature of speech-processing tasks, we could expand the coverage of modules to tasks challenging for current speech foundation models~\cite{dunbar2021zero, huang2024zero}. This will further enhance the power and adaptability of Speech-Copilot.
% Future research could explore the use of multiple modules or foundation models for each function and the application of reinforcement learning from human feedback~\cite{ouyang2022training} for optimal module selection.

% Comment out for camera-ready
\section{Acknowledgement}
This work was inspired by a discussion between Prof. Thomas Hain and Prof. Hung-yi Lee during a taxi ride at ASRU 2023. In addition, we thank the National Center for High-performance Computing (NCHC) of the National Applied Research Laboratories (NARLabs) in Taiwan for providing computational and storage resources.

% Speech-Copilot offers a practical and efficient approach to handling diverse, instruction-oriented speech-processing tasks. 
% By breaking down complex instructions into manageable sub-tasks and utilizing a flexible LLM-based agent, our framework reduces the effort needed for toolset construction and enhances task performance. 
% This work demonstrates the potential of our modular framework, validated by high marks on the Dynamic-SUPERB benchmark, and sets a foundation for future developments in speech processing.

% In the future, we can explore having multiple modules or foundation models to choose from for each corresponding function. 
% In this case, we should consider how to best select modules, such as using reinforcement learning from human feedback (RLHF)~\cite{ouyang2022training} to guide the model in choosing which modules to use.

\bibliographystyle{IEEEbib}
\bibliography{refs}

\end{document}